\documentclass[12pt]{article}
\usepackage{amssymb}
\usepackage{amsmath}
\def\theequation{\thesection.\arabic{equation}}

\newtheorem{theorem}{Theorem}

\def\thetheorem{\thesection.\arabic{theorem}}

\def\theprop{\thesubsection.\arabic{prop}}
\newtheorem{lemma}[theorem]{Lemma}
\def\thelemma{\thesection.\arabic{lemma}}

\def\thecor{\thesection.\arabic{cor}}

\def\theexam{\thesection.\arabic{exam}}

\def\theremark{\thesection.\arabic{remark}}
\newcommand{\eqa}{\begin{eqnarray}}
\newcommand{\eeqa}{\end{eqnarray}}
\newcommand{\beq}{\begin{equation}}
\newcommand{\eeq}{\end{equation}}
\newcommand{\nn}{\nonumber}

\newcommand{\pal}{\partial}

\newcommand{\R}{{\mathbb R}}

\newcommand{\pf}{\noindent{\it Proof \ }}

\title{On Hamiltonian perturbations of hyperbolic systems of conservation laws, II: universality of critical behaviour}
\author{{Boris Dubrovin}\\
{\small  \ SISSA, Via Beirut 2--4, 34014 Trieste, Italy}}
\begin{document}
\maketitle
\begin{abstract} Hamiltonian perturbations of the simplest hyperbolic equation $u_t + a(u) u_x=0$ are studied. We argue that the behaviour of solutions to the perturbed equation near the point of gradient catastrophe of the unperturbed one should be essentially independent on the choice of generic perturbation neither on the choice of generic solution. Moreover, this behaviour is described by a special solution to an integrable fourth order ODE.
\end{abstract}

\setcounter{equation}{0}
\setcounter{theorem}{0}
\section{Introduction}

In the present work we continue the study of Hamiltonian perturbations of hyperbolic PDEs initiated by the paper \cite{dlz}. We consider here the simplest case of a single equation in one spatial dimension
\beq\label{eq1}
u_t+a(u) u_x + \epsilon\left[ b_1(u) u_{xx} + b_2(u) u_x^2\right]
+\epsilon^2\left[b_3(u) u_{xxx} + b_4(u) u_x u_{xx} + b_5(u) u_x^3\right]+\dots=0.
\eeq
Here $\epsilon$ is a small parameter; the coefficient of $\epsilon^k$
is a graded homogeneous polynomial in the derivatives $u_x$, $u_{xx}$, \dots of the total degree $(k+1)$,
$$
\deg u^{(n)} =n, \quad n>0.
$$
The unperturbed equation
\beq\label{eq0}
u_t+a(u) u_x=0
\eeq
can be considered as the simplest example of a nonlinear hyperbolic system;
the smooth functions $b_1(u)$, $b_2(u)$ etc. determine the structure of the perturbation.

Such expansions arise, e.g., in the study of the long wave (also called dispersionless) approximations of evolutionary PDEs; see section \ref{exam} below for other mechanisms that yield perturbed equations of the form (\ref{eq1}).

The unperturbed equation (\ref{eq0}) admits a Hamiltonian description
of the form
\eqa\label{ham1}
&&
u_t +\{ u(x), H_0\}\equiv u_t +\pal_x\frac{\delta H_0}{\delta u(x)} =0
\\
&&
\nn\\
&&
\quad H_0 =\int f(u)\, dx, \quad f''(u) = a(u)
\nn\\
&&
\nn\\
&&\label{pb1}
\{ u(x) , u(y)\}=\delta'(x-y)
\eeqa

The perturbed equations of the form (\ref{eq1}) are considered up to equivalencies defined by {\it Miura-type transformations} \cite{DZ1} of the form
\beq\label{miura}
u\mapsto u+\sum_{k\geq 1} \epsilon^kF_k(u; u_x, \dots, u^{(k)})
\eeq
where $F_k(u; u_x, \dots, u^{(k)})$ is a graded homogeneous polynomial in the derivatives $u_x$, $u_{xx}$, \dots of the degree
$$
\deg F_k = k.
$$
Using results of \cite{get} (see also \cite{magri2, DZ1}) one can show that any Hamiltonian perturbation of the equation (\ref{eq0}) can be reduced to the form
\eqa\label{eq-ham}
&&
u_t +\pal_x\frac{\delta H}{\delta u(x)}=0, \quad H=H_0+\epsilon\, H_1 +\epsilon^2 H_2 +\dots
\nn\\
&&
\\
&&
H_k =\int h_k(u; u_x,\dots, u^{(k)})\, dx, \quad \deg h_k(u; u_x, \dots, u^{(k)})=k.
\nn
\eeqa
Recall that for $H=\int h(u; u_x, u_{xx}, \dots)\, dx$
$$
\frac{\delta H}{\delta u(x)}={\mathcal E} \, h
$$
where
$$
{\mathcal E}=\frac{\pal}{\pal u} -\pal_x \frac{\pal}{\pal u_x} +\pal_x^2 
\frac{\pal}{\pal u_{xx}}-\dots
$$ 
is the Euler - Lagrange operator. The following well known property of the Euler - Lagrange operator will be often used in this paper: ${\mathcal E} \, h=0$ {\it iff} there exists $h_1=h_1(u; u_x, \dots)$ such that $h=\mbox{const}+\pal_x h_1$. Note that we do not specify here the class of functions $u(x)$.
The Hamiltonians $H=H[u]$ can be ill defined (e.g., a divergent integral) but the evolutionary PDE (\ref{eq-ham}) makes sense. The crucial point for the subsequent considerations is the following statement (see, e.g., \cite{dick}): for two commuting Hamiltonians
$$
\{ H, F\}=0 \quad \Leftrightarrow \quad {\mathcal E} \left(
\frac{\delta H}{\delta u(x)} \, \pal_x \frac{\delta F}{\delta u(x)}\right) =0
$$
the evolutionary PDEs
$$
u_t +\pal_x\frac{\delta H}{\delta u(x)}=0 \quad\mbox{and}\quad u_s +\pal_x\frac{\delta F}{\delta u(x)}=0
$$
commute,
$$
(u_t)_s = (u_s)_t.
$$

For sufficiently small $\epsilon$ one expects to see no major differences in the behaviour of solutions to the perturbed and unperturbed equations (\ref{eq1}) and (\ref{eq0}) within the regions where the $x$-derivatives
are bounded. However the differences become quite serious near the critical point (also called the point of {\it gradient catastrophe}) where the derivatives of solution to the unperturbed equation tend to infinity.

Although the case of small viscosity perturbations has been well studied and understood (see \cite{bressan} and references therein), the critical behaviour of solutions to {\it general conservative perturbations}
(\ref{eq-ham}) to our best knowledge has not been investigated
(see the papers \cite{el, gm, gp, hl, ks, ll, llv, Lo} for the study of various particular cases).

The main goal of this paper is to formulate the Universality Conjecture
about the behaviour of a generic solution to the general  perturbed Hamiltonian equation near the point of gradient catastrophe of the unperturbed solution. We argue that, up to shifts, Galilean transformations and rescalings this behaviour essentially {\it does not} depend on the choice of solution neither on the choice of the equation (provided certain genericity assumptions hold valid). Moreover, this behaviour near the point $(x_0, t_0, u_0)$ is given by
\beq\label{univer}
u\simeq u_0 +a\,\epsilon^{2/7} U \left( b\, \epsilon^{-6/7} (x- a_0 (t-t_0)- x_0); c\, \epsilon^{-4/7} (t-t_0)\right) +O\left( \epsilon^{4/7}\right)
\eeq
where $U=U(X; T)$ is the unique real {\it smooth for all $X\in \R$} solution to the fourth order ODE
\beq\label{ode0}
X=T\, U -\left[ \frac16 U^3 +\frac1{24} ( {U'}^2 + 2 U\, U'' ) +\frac1{240} U^{IV}\right], \quad U'=\frac{dU}{dX} \quad \mbox{ etc.}
\eeq
depending on the parameter $T$. Here $a$, $b$, $c$ are some constants that depend on the choice of the equation and the solution, $a_0=a(v_0)$.

The equation (\ref{ode0}) appeared in \cite{bmp} (for the particular value  of the parameter $T=0$) in the study of the double scaling limit for the matrix model with the multicritical index $m=3$. It was observed that generic solutions to (\ref{ode0}) blow up at some point of real line; the conjecture about existence of a unique smooth solution has been formulated. To our best knowledge, this conjecture
remains open, although there are some supporting evidences   \cite{kapaev}.

The present paper is organized as follows. In Section 2 we classify all Hamiltonian perturbations up to the order $\epsilon^4$. They are parametrized by two arbitrary functions $c(u)$, $p(u)$. For the simplest example the perturbations of the Riemann wave equation $u_t+u\, u_x=0$ read
\eqa\label{riem2}
&&
u_t +u\, u_x + \frac{\epsilon^2}{24} \left[ 2 c\, u_{xxx} + 4 c' u_x u_{xx}
+ c'' u_x^3\right]+\epsilon^4 \left[ 2 p\, u_{xxxxx} \right.
\nn\\
&&
\\
&&\left.
+2 p'( 5 u_{xx} u_{xxx} + 3 u_x u_{xxxx}) + p''( 7 u_x u_{xx}^2 + 6 u_x^2 u_{xxx} ) +2 p''' u_x^3 u_{xx}\right]=0.
\nn
\eeqa
For $c(u)=\mbox{const}$, $p(u)=0$ this is nothing but the Korteweg - de Vries (KdV) equation; for other choices of the functions $c(u)$, $p(u)$ it seems not to be an integrable PDE. Remarkably, for arbitrary choice of the functional parameters the perturbed equation possesses an infinite family of 
{\it approximate symmetries} (see \cite{bgi, DZ1, km, str} for discussion of approximate symmetries). In principle our approach can be applied to classifying the Hamiltonian perturbations of higher orders. However, higher order terms do not affect the type of critical behaviour.
 
In Section 3
we establish an important property of {\it quasitriviality} of all perturbations (cf. \cite{DZ1, dlz, LZ2}). The quasitriviality is given by a substitution
\beq\label{quasi0}
u\mapsto u +\epsilon^2 K_2(u; u_x, u_{xx}, u_{xxx}) +\epsilon^4 K_4(u; u_x, \dots, u^{(6)})
\eeq
that transforms, modulo $O(\epsilon^6)$ the unperturbed equation (\ref{eq0}) to (\ref{eq-ham}). Here the functions $K_2$ and $K_4$ depend {\it rationally} on the $x$-derivatives. We also formulate the first part of our Main Conjecture that says that, for sufficiently small
$\epsilon$ the solution to the perturbed system exists at least on the same domain of the $(x,t)$-plane where the unperturbed solution is defined. In Section 4 we briefly discuss existence of a bihamiltonian structure compatible with the perturbation (see also Appendix below).
Some examples of perturbed Hamiltonian equations are described in Section 5. In Section 6 we recollect some properties of the ODE
(\ref{ode0}). Finally, in Section 7 we give the precise formulation of the Universality Conjecture and give some evidences supporting it. Because of lack of space we do not consider the numerical evidences supporting the idea of Universality; they will be given in a subsequent publication (see also \cite{gk}). In the last section we outline the programme of further researches towards
understanding of universality phenomena of critical behaviour in general  Hamiltonian perturbations of hyperbolic systems.
\vskip 0.5truecm
\noindent{\bf Acknowledgments.} This work is
partially supported by European Science Foundation Programme ``Methods of
Integrable Systems, Geometry, Applied Mathematics" (MISGAM), Marie Curie RTN ``European Network in Geometry, Mathematical Physics and Applications"  (ENIGMA), 
and by Italian Ministry of Universities and Researches (MIUR) research grant PRIN 2004
``Geometric methods in the theory of nonlinear waves and their applications".

\setcounter{equation}{0}
\setcounter{theorem}{0}
\section{Hamiltonian perturbations of the Riemann wave equation}

Let us start with the simplest case of Hamiltonian perturbations of the equation
\eqa\label{riem0}
&&
v_t+v\, v_x=0 \quad \Leftrightarrow \quad v_t +\{ v(x), H_0\}=0
\\
&&
\nn\\
&&
\{ v(x), v(y)\}=\delta'(x-y) 
\nn\\
&&
\nn\\
&&
\quad H_0 =\int \frac{v^3}6\, dx
\nn
\eeqa
\begin{lemma} \label{lemma1} Up to the order $O(\epsilon^4)$, all Hamiltonian perturbations of (\ref{riem0}) can be reduced to the form
\eqa\label{riem1}
&&
u_t +\pal_x \frac{\delta H}{\delta u(x)}=0
\nn\\
&&
H=\int \left[ \frac{u^3}6 - \epsilon^2 \frac{c(u)}{24} u_x^2
+\epsilon^4 \left( p(u) u_{xx}^2 + s(u) u_x^4\right)\right]\, dx
\eeqa
where $c(u)$, $p(u)$, $s(u)$ are arbitrary functions. Moreover, the function $s(u)$ can be eliminated by a Miura-type transform.
\end{lemma}

\pf The Hamiltonian must have the form
$$
H=H_0 +\epsilon H_1 + \dots +\epsilon^4 H_4
$$
where the density of $H_k$ is a graded homogeneous polynomial of the degree $k$. So, the density of $H_1$ is a total derivative:
$$
H_1=\int\alpha(u) u_x\, dx, \quad\alpha(u) u_x =\pal_x A(u), \quad A'(u) =\alpha(u).
$$
The density of the Hamiltonian $H_2$ modulo total derivatives must have the form
$$
-\frac{c(u)}{24} u_x^2
$$
for some function $c(u)$. Similarly,  $H_3$ must have the form 
$$
H_3 = \int c_1(u) u_x^3\, dx.
$$
Here $c_1(u)$ is another arbitrary function.

Let us show that $H_3$ can be eliminated by a Miura-type transform. Let us look for it in the form
\beq\label{canon}
u\mapsto u +\epsilon\{ u(x), F\} +\frac{\epsilon^2}2 \{\{u(x), F\}, F\} +\dots
\eeq
choosing
$$
F=\epsilon^2 \int \alpha(u) u_x^2\, dx.
$$
Such a transformation {preserves} the Poisson bracket. The change of the Hamiltonian $H$ will be given by
$$
\delta H = \epsilon\, \{ F, H\}+O(\epsilon^4).
$$
At the order $\epsilon^3$ one has
{\small$$
\delta H=\epsilon^3  \int \left[ \frac12 \alpha'(u) u_x^2 -\partial_x \left( \alpha\, u_x\right) \right] \, u\, u_x\, dx
=\frac{\epsilon^3}2\int \alpha(u) u_x^3\, dx.
$$}
So, choosing
$
\alpha(u)= - 2 c_1(u)
$
we kill the terms cubic in $\epsilon$. 

The rest of the proof is obvious: in order 4 all the Hamiltonians have the form
$$
H_4 =\int [ p(u) u_{xx}^2 + s(u) u_x^4]\, dx
$$
for some functions $p(u)$, $s(u)$. The last term can be killed by the canonical transformation of the form (\ref{canon})
generated by the Hamiltonian
$$
F=-\frac{\epsilon^3}2 \int s(u) u_x^3\, dx.
$$
The lemma is proved.

Choosing $s(u)=0$ one obtains the family (\ref{riem2}) of Hamiltonian perturbations of the Riemann wave equation 
depending on two arbitrary functions $c=c(u)$, $p=p(u)$.

We will now compare the symmetries of (\ref{riem0}) and those of the perturbed system (\ref{riem1}).
It is easy to see that the Hamiltonian equation
\eqa\label{volna0}
&&
v_s+a(v)v_x=0 \quad \Leftrightarrow v_s+ \{ v(x), H_f^0\} =0
\\
&&
H_f^0 =\int f(v)\, dx, \quad f''(v) =a(v)
\nn
\eeqa
is a symmetry of (\ref{riem0})  for any $a(v)$,
$$
(v_t)_s=(v_s)_t.
$$
Moreover, the Hamiltonians $H_f^0$ commute pairwise,
$$
\{ H_f^0, H_g^0\}=0 \quad \forall f=f(u), \quad \forall g=g(u).
$$
This family of commuting Hamiltonians is {\it complete} in the following sense. 
\begin{lemma}\label{maximal} The family of commuting Hamiltonians $H_f^0$ is {\rm maximal}, i.e., if $H=\int h(u; u_x, u_{xx}, \dots)\, dx$ commutes with all functionals of the form $H_f^0$ then
$$
h(u; u_x, u_{xx}, \dots) = g(u) +\pal_x (\dots)
$$
for some function $g(u)$.
\end{lemma}

We will now construct a perturbation of the Hamiltonians $H_f^0$ preserving the commutativity modulo $O(\epsilon^6)$. Like in Lemma \ref{lemma1} one can easily check that all the perturbations up to the order 4 must have the form
$$
H_f = \int \left\{ f(u) -\epsilon^2\frac{c_f(u)}{24}u_x^2 +\epsilon^4\left[ p_f(u) u_{xx}^2 +s_f(u) u_x^4\right]\right\}\, dx
$$
for some functions $c_f(u)$, $p_f(u)$, $s_f(u)$. To ensure commutativity one has to choose these functions as follows.

\begin{lemma}\label{lemma2} For any $f=f(u)$ the Hamiltonian flow
\eqa\label{volna1}
&&
u_s +\pal_x\frac{\delta H_f}{\delta u(x)}=0, \quad H_f =\int h_f\, dx
\nn\\
&&
h_f = f-\frac{\epsilon^2}{24} c\, f''' u_x^2+\epsilon^4\left[\left( p\,f''' + 
     \frac{{c}^2\,f^{(4)}}{480} \right) \,u_{xx}^2\right.
\\
&&\left.  - 
 \left(   
      \frac{c\,c''\,f^{(4)}}{1152} + \frac{c\,c'\,f^{(5)}}{1152} + 
     \frac{{c}^2\,f^{(6)}}{3456} +\frac{p'\,f^{(4)}}{6} +
     \frac{p\,f^{(5)}}{6}  -s\,f'''\right)\, u_{x}^4\right]
\nn
\eeqa
is a symmetry, modulo $O(\epsilon^6)$, of (\ref{riem1}). Moreover, the Hamiltonians $H_f$ commute pairwise:
$$
\{ H_f, H_g\} =O(\epsilon^6)
$$
for arbitrary two functions $f(u)$ and $g(u)$.
\end{lemma}

\pf One has to check the identity 
$$
{\mathcal E}\left(\frac{\delta H_f}{\delta u(x)} \pal_x \frac{\delta H_g}{\delta u(x)}\right)=0
$$
where ${\mathcal E}$
is the Euler - Lagrange operator. We leave this calculation as an exercise for the reader.

Observe that for $f=\frac{u^3}6$ the Hamiltonian $H_f$ coincides with (\ref{riem1}). Also for $f=u$ (the Casimir of the Poisson bracket) and
$f=\frac{u^2}2$ (the momentum) the perturbation is trivial,
$$
H_f= H_f^0.
$$

We do not know under what conditions on the functional parameters $c(u)$, $p(u)$ higher order perturbations can be added
to the Hamiltonians (\ref{volna1}) preserving the commutativity. The examples of Section \ref{exam} show that this can be done at least for some particular choices of the functions. However, the remark at the end of Section \ref{biha} suggests that the answer is not always affirmative.

\setcounter{equation}{0}
\setcounter{theorem}{0}
\section{Solutions to the perturbed equations. Quasitriviality}\label{qua}\par

Next question: existence of solutions to the perturbed equation for $t< t_C$. We will construct a {\it formal asymptotic solution} to (\ref{riem1}) (and also to all commuting flows (\ref{volna1})) valid on the entire interval $t<t_C$.
The basic idea: find a substitution 
$$
v\mapsto u= v +O(\epsilon)
$$
that transforms {\it all} solutions to {\it all unperturbed equations} of the form (\ref{volna0}) to solutions to the corresponding {\it perturbed equations} (\ref{volna1}).

{\bf Quasitriviality Theorem} {\it There exists a transformation
\beq\label{quasi}
v\mapsto u = v + \sum_{k=1}^4 \epsilon^k F_k(u; u_x, \dots, u^{(n_k)}),
\eeq
where $F_k$ are {\rm rational functions} in the derivatives homogeneous of the degree $k$, independent of $f=f(u)$,  that transforms all monotone solutions of (\ref{volna0}) to solutions, modulo $O(\epsilon^6)$, of (\ref{volna1}) and vice versa.}

The general quasitriviality theorem for evolutionary PDEs admitting a {\it bihamiltonian} description was obtained in \cite{dlz}\footnote{In a very recent paper \cite{LZ2} the quasitriviality result was proved, in all orders in $\epsilon$, for an {\it arbitrary} perturbation of the Riemann wave equation $v_t + v\, v_x=0$. It has also been  shown that the same transformation trivializes also all symmetries of the perturbed equation. }. As we do not assume {\sl a priori} existence of a bihamiltonian structure (see, however, the next section), we will give here a direct proof of quasitriviality for the family of commuting Hamiltonians (\ref{volna1}).

For convenience we chose 
$$
s(u)=\frac{c(u) \, c'''(u)}{3456}.
$$

\begin{theorem}\label{theorem2}  Introduce the following Hamiltonian
$$
K=\int \left[ \frac1{24}\epsilon\,c(u)\, u_x \,\log u_x 
+\epsilon^3 \left( \frac{c^2(u)}{5760} \frac{u_{xx}^3}{u_x^3} 
-\frac{p(u)}4 \frac{u_{xx}^2}{u_x} \right)\right]\, dx.
$$
Then the canonical transformation
$$
u\mapsto v =u +\epsilon \{ u(x), K\} +\frac{\epsilon^2}2 \{ \{ u(x), K\}, K\} +\dots
$$
satisfies
$$
H_f = \int f(v) \, dx +O(\epsilon^6) \quad \forall f(u).
$$
\end{theorem}

The inverse transformation is the needed quasitriviality. It is generated by the Hamiltonian
$$
-K = \int \left[ -\frac1{24}\epsilon\,c(v)\, v_x \,\log v_x 
-\epsilon^3 \left( \frac{c^2(v)}{5760} \frac{v_{xx}^3}{v_x^3} 
-\frac{p(v)}4 \frac{v_{xx}^2}{v_x} \right)\right]\, dx,
$$
that is
\eqa\label{quasi1}
&&
v\mapsto u=v -\epsilon\{ v(x),K \} +\frac{\epsilon^2}2 \{ \{ v(x), K\},K\}+\dots
\nn\\
&&
= v +\frac{\epsilon^2}{24} \partial_x \left( c \frac{v_{xx}}{v_x} 
+ c' v_x\right)
+\epsilon^4\partial_x\left[{c}^2\left(
\frac{v_{xx}^3}{360\,v_x^4} -\frac{7\, v_{xx} v_{xxx}}{1920\, v_x^3}
+\frac{v_{xxxx}}{1152\, v_x^2}
\right) _x
  \right.
\nn\\
&&+ 
 c\, c'\,\left( \frac{47\,{{v_{xx}}}^3}{5760\,{{v_{x}}}^3} - 
  \frac{37\,{v_{xx}}\,{v_{xxx}}}
   {2880\,{{v_{x}}}^2} + 
  \frac{5\,{v_{xxxx}}}{1152\,{v_{x}}}\right)+ {c'}^2\left(
  \frac{{v_{xxx}}}{384}- 
  \frac{{{v_{xx}}}^2}{5760\,{v_{x}}}\right) + c\,c''\left(
  \frac{{v_{xxx}}}{144}- 
  \frac{{{v_{xx}}}^2}{360\,{v_{x}}}\right)
\nn\\
&&  + 
  \frac1{1152}\left({7\,c'\,c''\,{v_{x}}\,{v_{xx}}} + 
  {c''}^2\,{{{v_{x}}}^3}+ 
  6\,c\,c'''\,{{v_{x}}\,{v_{xx}}}+ 
 c'\,c'''\, {{{v_{x}}}^3}+ 
  c\,c^{(4)}\,{{{v_{x}}}^3}\right)
\nn\\
&&\left.   
  + 
  p\,\left( \frac{{{v_{xx}}}^3}{2\,{{v_{x}}}^3} - 
  \frac{{v_{xx}}\,{v_{xxx}}}{{{v_{x}}}^2} + 
  \frac{{v_{xxxx}}}{2\,{v_{x}}} \right)+p' {v_{xxx}} + 
 p'' \frac{{v_{x}}\,{v_{xx}}}{2} 
\right]
\eeqa
In this formula $c=c(v)$, $p=p(v)$.

{\bf Main Conjecture, Part 1}. {\it  Let $v=v(x,t)$ be a smooth solution to the unperturbed equation $v_t+a(v)\, v_x=0$ defined for all $x\in\R$ and $0\leq t <t_0$ monotone in $x$ for any $t$. Then there exists a solution
$u=u(x, t; \epsilon)$ to the perturbed equation 
$$
u_t +\pal_x \frac{\delta H_f}{\delta u(x)} =0, \quad f''(u)=a(u)
$$
defined on the same domain in the $(x,t)$-plane
with the asymptotic at $\epsilon\to 0$ of the form (\ref{quasi1}). }

\setcounter{equation}{0}
\setcounter{theorem}{0}
\section{Are all Hamiltonian perturbations also bihamiltonian?}\label{biha}\par

All  unperturbed equations
$$
v_s + a(v)\, v_x=0
$$
are bihamiltonian w.r.t. the Poisson pencil (see the definition in \cite{DZ1})
\beq\label{biham}
\{ v(x), v(y)\}_1=\delta'(x-y), \quad \{ v(x), v(y)\}_2 = q(v(x))\delta'(x-y)+\frac12 q'(v) v_x \delta(x-y)
\eeq
for an arbitrary function $q(u)$,
\eqa
&&
v_s + \{ v(x), H_1\}_1 =v_s +\{ v(x), H_2\}_2=0, \quad H_1=\int f_1(v)\, dx , \quad H_2=\int f_2(v)\, dx
\nn\\
&&
f_1''(v) = a(v)=q(v) f_2''(v) +\frac12 q'(v) f_2'(v).
\nn
\eeqa
To show that (\ref{biham}) is a Poisson pencil it suffices to observe that the linear combination
\beq\label{pencil0}
\{ v(x), v(y)\}_2 -\lambda \, \{ v(x), v(y)\}_1=
\left(q(v(x)) -\lambda\right) \delta'(x-y) + \frac12 q'(v) v_x \delta(x-y)
\eeq
is the Poisson bracket associated \cite{dn83} with the flat metric
$$
ds^2 =\frac{dv^2}{q(v)-\lambda}.
$$

\begin{theorem} \label{biham-th} For $c(u)\neq 0$ the commuting Hamiltonians (\ref{volna1}) admit a unique bihamiltonian structure obtained by a deformation of
(\ref{biham}) with $q(u)$ satisfying
\beq\label{cons}
p(u) = \frac{c^2}{960} \left[ 5\,\frac{ c'}{c} - \frac{q''}{q'}\right], \quad s(u)=0.
\eeq
\end{theorem}

The proof of this result along with the explicit formula for the deformed bihamiltonian structure is sketched in the Appendix below.

The assumption $c\neq 0$ is essential: one can check that for $c(u)\equiv 0$ the Hamiltonians (\ref{volna1}) commute, modulo $O(\epsilon^6)$, {\it only} w.r.t. the standard Poisson bracket (\ref{pb1}). On the other side it turns out that for this particular choice of the functional parameters the deformation of commuting Hamiltonians {\it cannot} be extended to the order $O(\epsilon^8)$.

\setcounter{equation}{0}
\setcounter{theorem}{0}
\section{Examples}\label{exam}

{\bf Example 1.} For $c(u)=c_0=\mbox{const}$, $p(u)=s(u)=0$ one obtains  from (\ref{riem1}) the KdV equation
$$
u_t + u\, u_x +c_0\,\frac{\epsilon^2}{12} u_{xxx}=0.
$$
Choosing in (\ref{volna1}) 
$$
f(u) =\frac{u^{k+2}}{(k+2)!}
$$
one obtains the Hamiltonians of the KdV hierarchy
\eqa
&&
\frac{\pal u}{\pal t_k} + \pal_x \frac{\delta H_k}{\delta u(x)} =0, \quad 
H_k =\int h_k\, dx, \quad k\geq 0
\nn\\
&&
h_k  =\frac{u^{k+2}}{(k+2)!}-c_0\, \frac{\epsilon^2}{24} \frac{u^{k-1}}{(k-1)!} u_x^2 +c_0^2 \frac{\epsilon^4}{96} \left[ \frac{u^{k-2}}{5\, (k-2)!} u_{xx}^2 - \frac{u^{k-4}}{36\, (k-4)!} u_x^4\right]+O(\epsilon^6).
\nn
\eeqa
The quasitriviality transformation (\ref{quasi1}) takes the form \cite{bgi, DZ1}
\beq\label{quasi2}
v\mapsto u=v +\pal_x^2 \left[ \frac{\epsilon^2}{24} \, c_0\, \log v_x +{c_0}^2\epsilon^4\left(
\frac{v_{xx}^3}{360\,v_x^4} -\frac{7\, v_{xx} v_{xxx}}{1920\, v_x^3}
+\frac{v_{xxxx}}{1152\, v_x^2}
\right)\right] +O(\epsilon^6).
\eeq

\noindent{\bf Example 2.} The {\it Volterra lattice}
\begin{equation}\label{volt}
\dot q_n = q_n (q_{n+1} - q_{n-1})
\end{equation}
(also called difference KdV) has the following bihamiltonian structure \cite{ft}
\eqa\label{volt1}
&&
\{ q_n, q_m\}_1 = 2 q_n q_m (\delta_{n+1, m} - \delta_{n, m+1})
\\
&&
\nn\\
&&
\dot q_n =\{ q_n, H_1\}_1, \quad H_1 = \frac12 \sum \log q_n
\nn
\eeqa
\eqa\label{volt2}
&&
\{ q_n, q_m\}_2 =  q_n q_m\left\{ \left[ \frac{q_n+q_m}2 -2\right]\,
\left( \delta_{n, m+1} - \delta_{n,m-1}\right)
+\frac12 \delta_{n,m+2} -\frac12 \delta_{n,m-2}\right\}
\\
&&
\nn\\
&&
\dot q_n =\{ q_n, H_2\}_2, \quad H_2 = \sum q_n
\nn
\eeqa
After substitution
$$
q_n = e^{v(n\epsilon)}
$$
and division by $4 \epsilon$ one arrives at the following bihamiltonian structure
\beq\label{prima1}
\{ v(x), v(y)\}_1 =\frac1{4\epsilon} \left[ \delta(x-y+\epsilon) - \delta(x-y-\epsilon)\right]=\delta'(x-y) +\frac{\epsilon^2}3 \delta'''(x-y)+\dots
\eeq
\eqa\label{seconda1}
&&
\{ v(x), v(y)\}_2 = \left( 1-e^{v(x)}\right) \delta'(x-y) -\frac12 e^{v}v_x \delta(x-y)
\\
&&
+\epsilon^2 \left[ \frac1{12} (2-5\, e^v) \delta'''(x-y) -\frac58 e^v v_x \delta''(x-y)
\right.
\nn\\
&&\left.
-\frac38 e^v (v_{xx}+v_x^2) \delta'(x-y) -\frac1{12} e^v (v_{xxx} +3 v_x v_{xx} + v_x^3) \delta(x-y)\right] + O(\epsilon^4).
\nn
\eeqa
To compare this bihamiltonian structure with the one obtained in Theorem \ref{biham-th} the Poisson bracket (\ref{prima1}) must be reduced to the standard form
\beq\label{prima2}
\{ u(x), u(y\}_1 =\delta'(x-y)
\eeq
by means of the transformation
$$
u= \sqrt{\frac{\epsilon \pal_x}{\sinh{\epsilon \pal_x}}}\, v= v - \frac{\epsilon^2}{12}v_{xx} +\frac{\epsilon^4}{160} v_{xxxx}+O(\epsilon^6) .
$$
After the transformation the second bracket takes the form
\eqa\label{seconda2}
&&
\{ u(x), u(y)\}_2=\left( 1-e^{u(x)}\right) \delta'(x-y) -\frac12 e^{u}u_x \delta(x-y)
\\
&&
-\epsilon^2 e^{u(x)}\left[\frac14 \delta'''(x-y) + \frac38 u_x \delta''(x-y)
+\frac1{24} (7u_{xx} + 5 u_x^2) \delta'(x-y)
\right.
\nn\\
&&
\left. +\frac1{24}
( 2u_{xxx} + 4 u_x u_{xx} + u_x^3) \delta(x-y)\right] +O(\epsilon^4)
\nn
\eeqa
We leave as an exercise for the reader to compute the terms of order $\epsilon^4$ and to verify that the Poisson bracket (\ref{seconda2})
is associated with the functional parameters chosen as follows
$$
c(u)=2, \quad p(u) =-\frac1{240}, \quad q(u) = 1-e^u, \quad s(u) =\frac1{4320}.
$$

\noindent{\bf Example 3}.
The {\it Camassa - Holm equation} \cite{CH} (see also
\cite{Fo})
\beq\label{cam-holm}
v_t - \epsilon^2 v_{xxt} = \frac32 v\, v_x - \epsilon^2 \left[
 v_x v_{xx}+\frac12 v\, v_{xxx}\right]
\eeq
admits a bihamiltonian
description (cf. \cite{khesin-misl}) after doing the following Miura-type
transformation
\beq\label{cam-holm1}
u=v-\epsilon^2 v_{xx}.
\eeq
The bihamiltonian structure reads
\beq\label{cam-holm-pb1}
\{ u(x), u(y)\}_1 =\delta'(x-y) -\epsilon^2 \delta'''(x-y)
\eeq
\beq\label{cam-holm-pb2}
\{u(x), u(y)\}_2 =u(x) \delta'(x-y)+\frac12 u_x \delta(x-y)
.
\eeq

The Casimir $H_{-1}$ of the first Poisson bracket analytic in $\epsilon$
has the
form
$$
H_{-1} = \int h_{-1}dx, ~~h_{-1} =u(x).
$$
Applying the bihamiltonian recursion procedure one obtains
a sequence of commuting Hamiltonians $H_k=\int h_k dx$ of the hierarchy,
$$
h_0 = \frac12 u\, v, ~~ h_1 = \frac18 [ v^3 + u\, v^2 ], \dots
$$
The corresponding Hamiltonian flows
$$
u_{t_k}=\{u(x), H_k\}_1 \equiv (1-\epsilon^2 \pal_x^2) \pal_x \frac{\delta H_k}
{\delta u(x)}
$$
read
$$
u_{t_0} =u_x, ~~u_{t_1} = \frac32 v\, v_x - \epsilon^2 \left[ v_x v_{xx} +\frac12 v\, v_{xxx}\right], \dots.
$$
The last equation reduces to (\ref{cam-holm}) after the substitution
(\ref{cam-holm1}).

To compare the commuting Hamiltonians with those given in (\ref{volna1}) one must first reduce the first Poisson bracket to the standard form $\{ \tilde u(x), \tilde u( y)\}_1=\delta'(x-y)$ by the transformation
$$
\tilde u =\left( 1-\epsilon^2 \pal_x^2\right)^{-1/2} u=u+\frac12 \epsilon^2 u_{xx} +\frac38 \epsilon^4 u_{xxxx}+\dots.
$$
After the transformation the Camassa - Holm equation will read
$$
\tilde u_t =\frac32 \tilde u\, \tilde u_x +\epsilon^2 ( 2\tilde u_{x} \tilde u_{xx} + \tilde u\, \tilde u_{xxx}) +\epsilon^4( 5\, \tilde u_{xx} \tilde u_{xxx} + 3\, \tilde u_x \tilde u_{xxxx} +\tilde u\, \tilde u_{xxxxx})+\dots .
$$
It is easy to see that the commuting Hamiltonians of Camassa - Holm hierarchy are obtained from (\ref{volna1}) by the specialization
$$
c(u) = 8\, u, \quad p(u) =\frac{u}3,\quad q(u)=u, \quad s(u)=0.
$$

\setcounter{equation}{0}
\setcounter{theorem}{0}
\section{Introducing a special function}\par

Let us remind some properties of the differential equation 
\beq\label{main}
X=T\, U - \left[ \frac16 U^3 +\frac1{24} ( {U'}^2 + 2 U\, U'' ) +\frac1{240} U^{IV}\right]
\eeq
often considered as a 4th order analogue of the classical Painlev\'e-I equation.
First, it can be interpreted as a monodromy preserving deformation
of the following linear differential operator with polynomial coefficients
\beq\label{iso}
\frac{\pal\psi}{\pal z}= {\bf W}\, \psi
\eeq 
where the matrix ${\bf W}$ reads
\eqa\label{w}
&&
{\bf W}=-\frac{1}{120}\left(
\begin{array}{cc}
 12 {U} {U'}+8 z {U'}+{U'''} & 2(16 z^2+8 z \
{U}+6 {U}^2+ {U''}-60{ T}) \\
 & \\
 2\,w_{21} & -12 \
{U} {U'}-8 z {U'}-{U'''}
\end{array}
\right)
\nn\\
&&
\nn\\
&&
\mbox{where}
\nn\\
&&
w_{21}=32 \
z^3-16 z^2 {U}-2 z (2 {U}^2+ {U''}+60\, T )+8 {U}^3+2 {U''} {U}- {U'}^2+120{ X}
\nn
\eeqa
Indeed, it coincides with the compatibility conditions
$$
{\bf W}_X -{\bf U}_z +[{\bf W},{\bf U}]=0
$$
of the linear system (\ref{iso}) with
\beq\label{u}
\frac{\pal\psi}{\pal X} ={\bf U}\, \psi,
\quad
{\bf U}=\left(
\begin{array}{cc}
 0 & -1 \\
  & \\
 2 {U}-2 z & 0
\end{array}
\right)
\eeq
Moreover, the dependence of (\ref{iso}) on $T$ is isomonodromic {\it iff} the function $U(X)$ depends also on the parameter $T$ according to the KdV equation
\beq\label{kdv0}
U_T + U\, U' +\frac1{12} U'''=0.
\eeq
This is the spelling of the compatibility condition of the  linear system (\ref{iso}), (\ref{u}) with 
\beq\label{v}
\frac{\pal\psi}{\pal T} ={\bf V}\, \psi, \quad
{\bf V}=\frac16\,\left(
\begin{array}{cl}
 {U'} & 2 {U}+4 z \\
  & \\
8 z^2-4 z {U} -4 {U}^2-{U''} & -{U'}
\end{array}
\right)
\eeq
The {\it Painlev\'e property} readily follows from the isomonodromicity: singularities in the complex $(X,T)$-plane of general solution to (\ref{main}), (\ref{kdv0}) are poles \cite{kapaev}.

{\bf Main Conjecture, Part 2}. {\it The ODE (\ref{main}) has unique solution
$U=U(X;T)$ { smooth for all real {$X\in\R$} }for all real values of the parameter $T$.}

Note that, due to the uniqueness the solution in question satisfies the KdV equation (\ref{kdv0}).

For $T<< 0$ the solution of interest is very close to the unique root of the cubic equation
$$
X\simeq T\, U -\frac{U^3}6,
$$
that is,
\eqa
&&
U\simeq (-T)^{1/2}\left[ w + (-T)^{-7/2} \frac{3 w^2 -2}{3\, (w^2+2)^4}
\right.
\nn\\
&&
\nn\\
&&
\left.-(-T)^{-7}w\,\frac{189 w^4 -972 w^2 + 436}{9\, (w^2 + 2)^9}
 +O\left( (-T)^{-21/2}\right)\right]
 \nn\\
 &&
 \nn\\
 &&
 X=-(-T)^{3/2} \left( w+\frac16 w^3\right).
 \label{resh}
\eeqa
Same is true for any $T$ for $|X|>>0$.
For $T>>0$ the solution develops oscillations typical for dispersive waves within a region around the origin; one can use Whitham method to approximate $U(X;T)$ by modulated elliptic functions within the oscillatory zone  \cite{gp, pote}.
Thus the solution in question interpolates between the two types of asymptotic behaviour (cf. \cite{ks} where the role of the special solution $U(X;T)$ in the KdV theory was discussed).  

The solutions to the fourth order ODE (\ref{main}) can be parametrized \cite{kapaev} by the monodromy data (i.e., the collection of Stokes multipliers) of the linear differential operator
(\ref{w}) with coefficients polynomial in $z$. The solution corresponding to given Stokes multipliers can be reconstructed by solving certain Riemann - Hilbert problem. The particular values of the Stokes multipliers associated with the smooth solution in question have been conjectured in \cite{kapaev}.

\setcounter{equation}{0}
\setcounter{theorem}{0}
\section{Local Galilean symmetry  and critical behaviour}\par

We will now proceed to discussing the universality problem. Consider the perturbed PDE
\beq\label{pde-1}
u_t + \{ u(x), H_{ f}\}= u_t + a(u) u_x +O(\epsilon^2)=0, \quad  f''(u) =a(u).
\eeq

Let us apply the transformation (\ref{quasi1}) to the unperturbed solution $v=v(x,t)$ of 
\beq\label{non-p}
v_t +a(v) v_x=0
\eeq
obtained by the method of characteristics:
\beq\label{sol0}
x=a(v)\, t + b(v)
\eeq
for some smooth function $b(v)$. Let the solution arrive at the point of gradient catastrophe for some $x=x_0$, $t=t_0$, $v=v_0$. At this point one has
\eqa\label{cata}
&&
x_0=a(v_0) t_0 +b(v_0)
\nn\\
&&
0=a'(v_0) t_0 + b'(v_0)
\\
&&
0=a''(v_0) t_0 + b''(v_0)
\nn
\eeqa
(inflection point). Let us assume the following {\it genericity assumption}
\beq\label{alpha}
\kappa:= -(a'''(v_0) t_0 + b'''(v_0))\neq 0.
\eeq

Let us first remind the universality property for the critical behaviour of the unperturbed solutions: up to shifts, Galilean transformations and rescalings a generic solution to (\ref{non-p}) near $(x_0, t_0)$ behaves like the cubic root function. We will present this well known statement in the following form. Introduce the new variables 
\eqa
&&
\bar x = x-a_0 (t-t_0) -x_0
\nn\\
&&
\bar t= t-t_0
\nn\\
&&
\bar v = v-v_0.
\nn
\eeqa
 Let us do the following scaling transformation
\eqa\label{scala2}
&&
\bar x \mapsto \lambda\, \,\bar x
\nn\\
&&
\nn\\
&&
\bar t \mapsto \lambda^{\frac23} \,\bar t
\\
&&
\nn\\
&&
\bar v \mapsto \lambda^{\frac13}\, \bar v
\nn
\eeqa

\begin{lemma} After the rescaling (\ref{scala2}) any generic solution
to (\ref{non-p}) at the limit $\lambda\to 0$ for $t<t_0$ goes to the solution of the cubic equation
\beq\label{cubic}
\bar x = a_0' \bar v \,\bar t - \kappa\, \frac{\bar v^3}6.
\eeq
\end{lemma}

In these formulae
$a_0=a(v_0)$, $a_0'=a'(v_0)$. Note that the inequality
\beq\label{sign}
\kappa\, a_0' >0
\eeq
must hold true in order to have the solution well defined for $t<t_0$ near the point of generic gradient catastrophe (\ref{cata}).

To prove the lemma it suffices to observe that, after the rescaling (\ref{scala2}) and division by $\lambda$ the equation (\ref{sol0}) yields
$$
\bar x = a_0' \bar v \,\bar t - \kappa\, \frac{\bar v^3}6 +O\left( \lambda^{1/3}\right).
$$

The parameter $\kappa$ can be eliminated from (\ref{cubic}) by a rescaling. The resulting cubic  function can be interpreted as the universal unfolding of the $A_2$ singularity \cite{arnold}. Our basic observation we are going to explain now is that, after  a Hamiltonian perturbation the $A_2$ singularity transforms to the special solution of (\ref{ode0}) described above.

Let us look for a solution to the perturbed PDE (\ref{pde-1}) in the form of a formal power series
\beq\label{sol-form}
u=u(x,t; \epsilon) = v(x,t) +\sum_{k\geq 1} \epsilon^k v_k(x,t)
\eeq
with $v(x,t)$ given by (\ref{sol0})
satisfying (\ref{pde-1}) modulo $O(\epsilon^5)$. We will say that such a solution is {\it monotone} at the point $x=x_0$, $t=t_0$ if
$$
u_x(x_0, t_0; 0) \equiv v_x(x_0, t_0)\neq 0.
$$
According to the results of Section \ref{qua} all monotone solutions of the form (\ref{sol-form}) can be obtained by applying the transformation (\ref{quasi1}) to the nonperturbed solution (\ref{non-p}) (more precisely, one has to allow $\epsilon$-dependence of the function $b(u)$).

\begin{lemma} Let us perform the rescaling (\ref{scala2}) along with 
\beq\label{scala3}
\epsilon\mapsto\lambda^{7/6}\epsilon.
\eeq
in the quasitriviality transformation (\ref{quasi1}). Then the resulting solution to the perturbed PDE will be equal to
\beq\label{univer1}
u=v_0 +\lambda^{1/3} \left\{\bar v +\pal_x^2 \left[ \frac{\epsilon^2}{24} \, c_0\, \log \bar v_x +{c_0}^2\epsilon^4\left(
\frac{\bar v_{xx}^3}{360\,\bar v_x^4} -\frac{7\, \bar v_{xx} \bar v_{xxx}}{1920\, \bar v_x^3}
+\frac{\bar v_{xxxx}}{1152\, \bar v_x^2}
\right)\right]
\right\}+O\left( \lambda^{2/3}\right)
\eeq
(cf. (\ref{quasi2})) where
\beq\label{c0}
c_0=c(v_0),
\eeq
$\bar v=\bar v(x,t)$ is the solution to the cubic equation (\ref{cubic}).
\end{lemma}

Proof is straightforward.

It remains to identify (\ref{univer1}) with the formal asymptotic solution
(\ref{resh}) to the ODE (\ref{main}). This can be done by a direct substitution. An alternative way is to observe that, near the point of gradient catastrophe the perturbed PDE acquires an {\it additional Galilean symmetry}. Indeed, according to the previous lemma, locally one can replace the functions $c(u)$, $p(u)$ by constants
$c_0=c(v_0)$, $p_0=p(v_0)$ (the constant $p_0$, however, does not enter in the leading term of the asymptotic expansion in powers of $\lambda^{1/3}$). Let us show that in this situation any solution to the perturbed PDE of the form (\ref{sol-form}) satisfies also a fourth order ODE.

\begin{lemma} Let $c(u)=c_0$, $p(u)=p_0$. Then for any solution $u(x,t;\epsilon)$ of the form (\ref{sol-form}) monotone at the point $(x_0, t_0)$ there exists a formal series
$$
g(u;\epsilon) = g_0(u) +\sum_{k\geq 1} \epsilon^k g_k(u)
$$
such that  for arbitrary $x$, $t$ sufficiently close to $x_0$, $t_0$ the function $u(x, t;\epsilon)$ satisfies, modulo $O(\epsilon^5)$, the following fourth order ODE
\beq\label{order4}
x= t\,\frac{\delta H_{f'}}{\delta u(x)} +\frac{\delta H_{g'}}{\delta u(x)} .
\eeq
Here
$$
g''_0(u)=b(u).
$$
\end{lemma}

\pf It is easy to see that the flow
\beq\label{sym}
u_\tau =1-t\, \pal_x\frac{\delta H_{f'}}{\delta u(x)}
\eeq
is a symmetry of (\ref{pde-1}).
Combining this symmetry with one of the commuting flows
$$
u_s +\pal_x \frac{\delta H_{g'}}{\delta u(x)}=0
$$
one obtains another symmetry. The set of stationary points of this combination
$$
\pal_x\left( t\, \frac{\delta H_{f'}}{\delta u(x)}+ \frac{\delta H_{g'}}{\delta u(x)}-x\right)=0
$$
is therefore invariant for the $t$-flow. Considering the limit $\epsilon\to 0$ it is easy to see that the integration constant vanishes on the solution (\ref{quasi1}), (\ref{non-p}). The lemma is proved. 
 
The ODE for the function $u(x)$ is closely related to the so-called {\it string equation} known in matrix models and topological field theory (see, e.g., \cite{DZ1}).  Explicitly
{\small\eqa\label{string}
&&
x=t\, a(u) + b(u) +c_0\frac{\epsilon^2}{24}\left\{ t\, \left[ 2  \,a'' u_{xx} + a''' u_x^2\right] + \left[ 2 \, b'' u_{xx} + b''' u_x^2\right] \right\}
\\
&&
+\epsilon^4 \left\{ \left[  2p_0\, \left(t\,a'' +b''\right)+ \frac1{240} c_0^2 \left(t\, a'''+b'''\right) 
 \right] u_{xxxx}
\right.
\nn\\
&&
\left[ 4 \, p_0\,\left(t\, a'''+b'''\right)+\frac1{120} c_0^2 \left(t\, a^{IV}+b^{IV}\right) \right] u_{xxx} u_x
\nn\\
&&
+\left[ 4 p_0\, \left(  t\, a^{IV}+b^{IV}\right)   +\frac{11}{1440}
c_0^2 \left( t\, a^V+b^V\right) 
 \right] u_{xx} u_x^2
\nn\\
&&\left.
+\left[\frac12 p_0\,\left(t\,  a^V +b^V\right)
+\frac1{1152} c_0^2 \left( t\, a^{VI}+b^{VI}\right)
\right]\, u_x^4\right\}.
\nn
\eeqa}

Let us call the solution {\it generic} if, along with the condition $\kappa:=-( a'''(v_0) t_0 + b'''(v_0))\neq 0$ it also satisfies
\beq\label{gamma}
c_0:= c(v_0)\neq 0.
\eeq

{\bf Main Conjecture, Part 3}. 
{\it The generic solution described in the Main Conjecture, Part 1 can be extended up to $t=t_0+\delta$ for sufficiently small positive $\delta=\delta(\epsilon)$; near the point $(x_0, t_0)$ it behaves in the following way
\beq\label{universal}
u\simeq v_0 + \left(\frac{\epsilon^2 c_0}{\kappa^2}\right)^{1/7} U\left( \frac{x-a_0 (t-t_0)-x_0}{(\kappa \,c_0^3\,\epsilon^6)^{1/7}}; ~   \frac{a_0'(t-t_0)}{(\kappa^3 c_0^2\epsilon^{4})^{1/7}}\right) +O\left( \epsilon^{4/7}\right).
\eeq}

To arrive at the asymptotic formula (\ref{universal}) we do in (\ref{string}) the rescaling of the form (\ref{scala2}) along with (\ref{scala3}).
After substitution to the equation (\ref{string}) and division by $\lambda$, one obtains
\eqa
&&
\bar x =a_0' \bar u\, \bar t -\kappa \left[ \frac{\bar u^3}6 +\frac{\epsilon^2}{24} c_0\left( \bar u_x^2 + 2 \bar u \,\bar u_{xx}\right)
+\frac{\epsilon^4}{240} c^2_0\bar u_{xxxx}\right] +O\left( \lambda^{1/3}\right).
\nn
\eeqa
In derivation of this formula we use that the monomial of the form
$$
\epsilon^k  u_x^{i_1} u_{xx}^{i_2} u_{xxx}^{i_3}\dots$$
after the rescaling will be multiplied by $\lambda^D$ with
$$
D=\frac16 k +\frac13 (i_1+i_2+\dots)
$$
due to the degree condition
$$
i_1+2\, i_2 + 3\, i_3+\dots = k.
$$
Adding the terms of higher order $k>4$ will not change the leading term.
Choosing
$$
\lambda=\epsilon^{6/7}c_0^{3/7}
$$
we arrive at the needed asymptotic formula. 

Clearly the above arguments require existence and uniqueness of the solution to (\ref{ode0}) smooth on the real line described in the Main Conjecture, Part 2.

\setcounter{equation}{0}
\setcounter{theorem}{0}
\section{Concluding remarks}\par
We have presented arguments supporting the conjectural universality of critical behaviour of solutions to generic Hamiltonian perturbations of a hyperbolic equation of the form (\ref{eq0}).
In subsequent publications we will study the Main Conjecture in more details. The possibilities of using the idea of Universality in numerical algorithms to dealing with oscillatory behaviour of solutions to Hamiltonian PDEs will be explored.  We will also proceed to the study of singularities of generic solutions to integrable Hamiltonian hyperbolic systems of conservation laws
\beq\label{hyp-ham}
u^i_t +\pal_x \left( \eta^{ij}\frac{\pal h(u)}{\pal u^j}\right)=0, \quad \eta^{ji}=\eta^{ij}, \quad \det(\eta^{ij})\neq 0.
\eeq
Recall that, according to the results of \cite{tsarev} the system (\ref{hyp-ham}) is integrable if it diagonalizes in a system of curvilinear
coordinates $v^k=v^k(u)$, $k=1, \dots, n$ for the Euclidean/pseudo-Euclidean metric 
$$
ds^2 =\eta_{ij} du^i du^j =\sum_{k=1}^n g_k(v) (dv^k)^2, \quad \left( \eta_{ij}\right):= \left( \eta^{ij}\right)^{-1},
$$
$$
v^k_t + \lambda^k(v) v^k_x =0, \quad k=1, \dots, n
$$
(in this formula no summation over repeated indices!). All Hamiltonian perturbations
of the hyperbolic system (\ref{hyp-ham}) can be written in the form
$$
u^i_t + \pal_x \left( \eta^{ij}\frac{\delta H}{\delta u^j(x)}\right)=0,\quad H=\int \left[ h(u) +\sum_{k\geq 1} \epsilon^k h_k(u; u_x, \dots, u^{(k)})\right]\, dx, \quad \deg h_k =k.
$$
We plan to study symmetries of the perturbed Hamiltonian hyperbolic systems. In particular, we will classify the perturbations preserving integrability and study the correspondence between the types of critical behaviour of the perturbed and unperturbed systems. The next step would be to extend our approach to Hamiltonian perturbations of spatially multidimensional hyperbolic systems (cf. \cite{dps}).

\def\thetheorem{A.\arabic{theorem}}
\def\theprop{A.\arabic{prop}}
\def\thelemma{A.\arabic{lemma}}
\def\thecor{A.\arabic{cor}}
\def\theexam{A.\arabic{exam}}
\def\theremark{A.\arabic{remark}}
\def\theequation{A.\arabic{equation}}

\setcounter{equation}{0}
\appendix
\makeatletter
\renewcommand{\@seccntformat}[1]{{Appendix:}\hspace{-2.3cm}}
\makeatother
\renewcommand{\thesection}{Appendix:}
\section{\quad\qquad \ \ Bihamiltonian structures associated with the perturbations of the Riemann wave hierarchy}

\begin{theorem}\label{theorem3}  For arbitrary two functions $c=c(u)$, $q=q(u)$ the family of Hamiltonians (\ref{volna1}) with 
\beq\label{cons1}
p(u) = \frac{c^2}{960} \left[ 5\,\frac{ c'}{c} -  \frac{q''}{q'}\right], \quad s(u)=0.
\eeq
is commutative
\beq\label{comm2}
\{ H_f, H_g\}_{1,2}=0\, \left({\rm mod} ~ O(\epsilon^6)\right)\quad \forall f=f(u), ~\forall g=g(u)
\eeq
with respect to the Poisson pencil of the form
$$
\{ u(x), u(y)\}_1 =\delta'(x-y),
$$
$$
\{ u(x), u(y)\}_2=\{ u(x), u(y)\}^{[0]}+\epsilon^2 \{ u(x), u(y)\}^{[2]}+\epsilon^4 \{ u(x), u(y)\}^{[4]} +O(\epsilon^6).
$$
Here the terms of order 0:
$$
\{ u(x), u(y)\}^{[0]}_2 =q(u)\delta'(x-y) +\frac12 q'(u) u_x \delta(x-y)
$$
All terms of higher orders are uniquely determined from the bicommutativity (\ref{comm2}) provided validity of the constraint (\ref{cons1}). Namely, the terms of order 2:
$$
\{ u(x), u(y)\}^{[2]}_2=
\frac{ {c}{q'} }{8}\delta'''(x-y)
 +\frac3{16}\left( 
   {{c}{q'}}\right)'{u_{x}}\delta''(x-y)
   $$
   $$+
 \left[\left(\frac{ {c''}{q'} \
 }{16} +\frac{{c'}{q''}}
    {6} + \frac{5{c}{q'''}}
    {48}\right){{u_{x}}}^2 + \frac{{c'}{q'}{u_{xx}}}{16} + 
   \frac{7{c}{q''}{u_{xx}}}{48}\right]\delta'(x-y)
   $$
   $$+
 \left[ \left(\frac{ {c''}{q''}  }
    {48} + \frac{{c'}{q'''}}{24} + 
   \frac{{c}{q^{(4)}}}{48}\right)\,{{u_{x}}}^3 + 
  \frac1{12}\left( {{c'}{q''}}
     + {{c}{q'''}}\right){u_{x}}{u_{xx}} + \frac{{c}{q''}
      }{24}{u_{xxx}}\right]\delta(x-y) 
$$
The terms of order 4: 
{\small$$
\{  u(x), u(y)\}^{[4]}_2=\frac1{192}\left(3{{c}{c'}{q'}} + 
   {{{c}}^2{q''}}\right) \delta^{V}(x-y)+
 \frac5{384}\left(3{{c}{c'}{q'}} + 
   {{{c}}^2{q''}}\right)'\,{u_{x}}\,\delta^{IV}(x-y)
      $$
      $$+
      \left[\left(\frac{3{c'}{c''}
      {q'}}{32} + 
   \frac{{c}{c'''}{q'}
      }{32} + 
   \frac{3{{c'}}^2{q''}}{32} + 
   \frac{5{c}{c''}{q''}
      }{48} - 
   \frac{{c}{c'}{{q''}}^2
      {{u_{x}}}^2}{240{q'}} + 
   \frac{{{c}}^2{{q''}}^3}
    {480{{q'}}^2} + 
   \frac{19{c}{c'}{q'''}
      }{192} - 
   \frac{3{{c}}^2{q''}{q'''}
      }{640{q'}} + 
   \frac{{{c}}^2{q^{(4)}}}{64}\right)\, {{u_{x}}}^2
   \right.
   $$
   $$\left. + 
   \left(\frac{3{{c'}}^2{q'}}{64} + 
   \frac{3{c}{c''}{q'}}
    {64} + \frac{17{c}{c'}{q''}
      }{192} - \frac{{{c}}^2
      {{q''}}^2}{480{q'}} + 
   \frac{19{{c}}^2{q'''}}{960}\right)\,{u_{xx}}\right]\delta'''(x-y)
   $$
   $$+
 \left[\left( \frac{3{{c''}}^2{q'}}{128} + 
   \frac{{c'}{c'''}{q'}
      }{32} + 
   \frac{{c}{c^{(4)}}{q'}
      }{128} + 
   \frac{19{c'}{c''}{q''}
      }{128} + 
   \frac{23{c}{c'''}{q''}
      }{384} + \frac{5{c}{c'}
      {q^{(4)}}}{64} + 
   \frac{7{c}{c''}{q'''}
      }{64} + 
   \frac{{{c}}^2{q^{(5)}}}{96}+ 
   \frac{3{{c'}}^2{q'''}}{32}
      \right.\right.
      $$
      $$\left.\left. - 
   \frac{{{c'}}^2{{q''}}^2}
    {160{q'}} - \frac{{c}{c''}
      {{q''}}^2}{160{q'}} + 
   \frac{{c}{c'}{{q''}}^3
      }{80{{q'}}^2} - 
   \frac{{{c}}^2{{q''}}^4}
    {160{{q'}}^3} - 
   \frac{17{c}{c'}{q''}
      {q'''}}{640{q'}} + 
   \frac{21{{c}}^2{{q''}}^2{q'''}
      }{1280{{q'}}^2} - 
   \frac{9{{c}}^2{{q'''}}^2}
    {1280{q'}} - 
   \frac{9{{c}}^2{q''}{q^{(4)}}
      }{1280{q'}}  \right)\,{{u_{x}}}^3
   \right.
   $$
   $$
   \left.+ 
  \left( \frac{9{c'}{c''}{q'}
     }{64} + 
   \frac{3{c}{c'''}{q'}
     }{64} + 
   \frac{11{{c'}}^2{q''}}{64} + \frac{13{c}{c''}
      {q''}}{64} - 
   \frac{3{c}{c'}{{q''}}^2
      }{160{q'}} + 
   \frac{3{{c}}^2{{q''}}^3}{320{{q'}}^2} + 
   \frac{69{c}{c'}{q'''}
   }{320} - 
   \frac{13{{c}}^2{q''}{q'''}
      }{640{q'}} + 
   \frac{3{{c}}^2{q^{(4)}}
      }{80} \right)\,  {u_{x}}{u_{xx}}
      \right.
      $$
      $$
      \left.+ \left(\frac{{{c'}}^2{q'}
      }{32} + \frac{{c}{c''}
      {q'}}{32} + 
   \frac{13{c}{c'}{q''}
      }{192} - \frac{{{c}}^2
      {{q''}}^2}{320{q'}} + 
   \frac{{{c}}^2{q'''}}{60}\right)\,{u_{xxx}}\right]\delta''(x-y)
   $$
   $$+
  \left[\left(\frac{{{c''}}^2{q''}}{48} + 
   \frac{{c'}{c'''}{q''}
     }{32} + 
   \frac{{c}{c^{(4)}}{q''}
     }{96} - 
   \frac{{c'}{c''}{{q''}}^2
     }{160{q'}} - 
   \frac{{c}{c'''}{{q''}}^2
     }{480{q'}} + 
   \frac{{{c'}}^2{{q''}}^3}
    {160{{q'}}^2} + 
   \frac{{c}{c''}{{q''}}^3
     }{160{{q'}}^2} - 
   \frac{{c}{c'}{{q''}}^4
     }{80{{q'}}^3} + 
   \frac{{{c}}^2{{q''}}^5}
    {160{{q'}}^4} + 
   \frac{35{c'}{c''}{q'''}
     }{384}
     \right.\right.
     $$
     $$ + 
   \frac{5{c}{c'''}{q'''}
     }{128} - 
   \frac{9{{c'}}^2{q''}{q'''}
     }{640{q'}} - 
   \frac{9{c}{c''}{q''}
      {q'''}}{640{q'}} + 
   \frac{11{c}{c'}{{q''}}^2
      {q'''}}{320{{q'}}^2} - 
   \frac{13{{c}}^2{{q''}}^3{q'''}
     }{640{{q'}}^3} - 
   \frac{{c}{c'}{{q'''}}^2
     }{64{q'}} + 
   \frac{19{{c}}^2{q''}{{q'''}}^2
     }{1280{{q'}}^2} + 
   \frac{17{{c'}}^2{q^{(4)}}}{384} 
   $$
   $$\left.\left.+ 
   \frac{5{c}{c''}{q^{(4)}}
     }{96} - 
   \frac{{c}{c'}{q''}{q^{(4)}}
     }{64{q'}} + 
   \frac{17{{c}}^2{{q''}}^2{q^{(4)}}
     }{1920{{q'}}^2} - 
   \frac{11{{c}}^2{q'''}{q^{(4)}}
     }{1280{q'}} + 
   \frac{35{c}{c'}{q^{(5)}}
     }{1152} - 
   \frac{11{{c}}^2{q''}{q^{(5)}}
     }{3840{q'}}+ 
   \frac{{{c}}^2{q^{(6)}}}{288}\right) \,{{u_{x}}}^4
      \right.
      $$
      $$\left.  + 
   \left(\frac{3{{c''}}^2{q'}}{128} + \frac{{c'}{c'''}
      {q'}}{32} + 
   \frac{{c}{c^{(4)}}{q'}
      }{128} + 
   \frac{91{c'}{c''}{q''}
      }{384} + 
   \frac{37{c}{c'''}{q''}
      }{384} - 
   \frac{{{c'}}^2{{q''}}^2}{60{q'}} - 
   \frac{{c}{c''}{{q''}}^2
      }{60{q'}} + 
   \frac{{c}{c'}{{q''}}^3
      }{30{{q'}}^2}- 
   \frac{{{c}}^2{{q''}}^4}{60{{q'}}^3} + 
   \frac{59{{c'}}^2{q'''}}{320}
      \right.\right.
      $$
      $$\left. \left.  + \frac{53{c}{c''}
      {q'''}}{240} - 
   \frac{47{c}{c'}{q''}
      {q'''}}{640
      {q'}} + \frac{173{{c}}^2
      {{q''}}^2{q'''}}{3840{{q'}}^2} - 
   \frac{77{{c}}^2{{q'''}}^2}{3840{q'}} + 
   \frac{169{c}{c'}{q^{(4)}}
      }{960} - 
   \frac{77{{c}}^2{q''}{q^{(4)}}
      }{3840{q'}} + 
   \frac{73{{c}}^2{q^{(5)}}}{2880}\right)\,{{u_{x}}}^2{u_{xx}}
      \right.
      $$
      $$\left.
       + \left(\frac{3{c'}{c''}
      {q'}}{128} + 
   \frac{{c}{c'''}{q'}
      }{128} + 
   \frac{5{{c'}}^2{q''}}{96} + 
   \frac{{c}{c''}{q''}
      }{16} - 
   \frac{{c}{c'}{{q''}}^2
      }{80{q'}} + 
   \frac{{{c}}^2{{q''}}^3}
    {160{{q'}}^2} + 
   \frac{157{c}{c'}{q'''}
      }{1920} - 
   \frac{5{{c}}^2{q''}{q'''}
      }{384{q'}} + 
   \frac{31{{c}}^2{q^{(4)}}}
    {1920}\right)\,{{u_{xx}}}^2
    \right.
    $$
    $$\left.
     +\left( \frac{3{c'}{c''}{q'}
      }{64} + 
   \frac{{c}{c'''}{q'}}{64} + \frac{{{c'}}^2{q''}
      }{12} + 
   \frac{3{c}{c''}{q''}
      }{32} - 
   \frac{{c}{c'}{{q''}}^2
      }{60{q'}} + 
   \frac{{{c}}^2{{q''}}^3}{120{{q'}}^2} + 
   \frac{19{c}{c'}{q'''}
      }{160} - 
   \frac{11{{c}}^2{q''}{q'''}
      }{640{q'}} + 
   \frac{11{{c}}^2{q^{(4)}}}{480} \right)\,{u_{x}}{u_{xxx}}
      \right.
      $$
      $$\left.
      + \left(\frac{{{c'}}^2{q'}
      }{128} + \frac{{c}{c''}
      {q'}}{128} + 
   \frac{11{c}{c'}{q''}
      }{384} - \frac{{{c}}^2
      {{q''}}^2}{320{q'}} + 
   \frac{17{{c}}^2{q'''}}{1920}\right)\,{u_{xxxx}}\right]\delta'(x-y)
   $$
   $$+
 \left[ \left(\frac{{{c''}}^2{q'''}}{192} + 
   \frac{{c'}{c'''}{q'''}}
    {128} + \frac{{c}{c^{(4)}}{q'''}
      }{384} - 
   \frac{{c'}{c''}{q''}{q'''}
      }{640{q'}} - 
   \frac{{c}{c'''}{q''}{q'''}
      }{1920{q'}} + 
   \frac{{{c'}}^2{{q''}}^2{q'''}
      }{640{{q'}}^2} + 
   \frac{{c}{c''}{{q''}}^2
      {q'''}}{640{{q'}}^2} - 
   \frac{{c}{c'}{{q''}}^3
      {q'''}}{320{{q'}}^3} 
    \right.\right.
      $$
      $$\left.\left. + 
   \frac{{{c}}^2{{q''}}^4{q'''}
      }{640{{q'}}^4} - 
   \frac{{{c'}}^2{{q'''}}^2}
    {640{q'}}
  - \frac{{c}{c''}
      {{q'''}}^2}{640{q'}} + 
   \frac{3{c}{c'}{q''}
      {{q'''}}^2}{640{{q'}}^2} - 
   \frac{{{c}}^2{{q''}}^2{{q'''}}^2
      }{320{{q'}}^3} + 
   \frac{{{c}}^2{{q'''}}^3}
    {1280{{q'}}^2}   + 
   \frac{7{c'}{c''}{q^{(4)}}
      }{384}    + 
   \frac{{c}{c'''}{q^{(4)}}}
    {128} - \frac{{{c'}}^2{q''}{q^{(4)}}
      }{640{q'}}   \right.\right.
      $$
      $$\left.\left.   - 
   \frac{{c}{c''}{q''}{q^{(4)}}
      }{640{q'}}  + 
   \frac{{c}{c'}{{q''}}^2
      {q^{(4)}}}{320{{q'}}^2} 
      - 
   \frac{{{c}}^2{{q''}}^3{q^{(4)}}
      }{640{{q'}}^3} - 
   \frac{3{c}{c'}{q'''}
      {q^{(4)}}}{640{q'}} + 
   \frac{13{{c}}^2{q''}{q'''}
      {q^{(4)}}}{3840{{q'}}^2} - 
   \frac{{{c}}^2{{q^{(4)}}}^2}
    {1280{q'}} + \frac{17{{c'}}^2{q^{(5)}}
      }{2304}       \right.\right.
      $$
      $$\left.\left. + 
   \frac{5{c}{c''}{q^{(5)}}
      }{576} - 
   \frac{{c}{c'}{q''}{q^{(5)}}
      }{640{q'}}
+ 
   \frac{{{c}}^2{{q''}}^2{q^{(5)}}
      }{1280{{q'}}^2} - 
   \frac{{{c}}^2{q'''}{q^{(5)}}
      }{960{q'}} + 
   \frac{5{c}{c'}{q^{(6)}}
      }{1152} - 
   \frac{{{c}}^2{q''}{q^{(6)}}
      }{3840{q'}} + 
   \frac{{{c}}^2{q^{(7)}}}{2304}\right)\,{{u_{x}}}^5
   \right.
   $$
   $$\left.
    + 
  \left( \frac{{{c''}}^2{q''}}{64} + \frac{{c'}{c'''}
      {q''}}{48} + 
   \frac{{c}{c^{(4)}}{q''}
      }{192} - 
   \frac{{c'}{c''}{{q''}}^2
      }{160{q'}} - 
   \frac{{c}{c'''}{{q''}}^2
      }{480{q'}} + 
   \frac{{{c'}}^2{{q''}}^3}{160{{q'}}^2} + 
   \frac{{c}{c''}{{q''}}^3
      }{160{{q'}}^2} - 
   \frac{{c}{c'}{{q''}}^4
      }{80{{q'}}^3} + 
   \frac{{{c}}^2{{q''}}^5}{160{{q'}}^4}
   \right.\right.
   $$
   $$\left.\left. + 
   \frac{97{c'}{c''}{q'''}
      }{960} + 
   \frac{13{c}{c'''}{q'''}
      }{320} - 
   \frac{{{c'}}^2{q''}{q'''}
      }{60{q'}} - 
   \frac{{c}{c''}{q''}{q'''}
      }{60{q'}} + 
   \frac{19{c}{c'}{{q''}}^2
      {q'''}}{480
      {{q'}}^2}- \frac{11{{c}}^2
      {{q''}}^3{q'''}}{480{{q'}}^3} - 
   \frac{{c}{c'}{{q'''}}^2
      }{48{q'}} + 
   \frac{3{{c}}^2{q''}{{q'''}}^2
      }{160{{q'}}^2}
      \right.\right.
      $$
      $$\left.\left.  + 
   \frac{19{{c'}}^2{q^{(4)}}}{320} + \frac{67{c}{c''}
      {q^{(4)}}}{960} - 
   \frac{{c}{c'}{q''}{q^{(4)}}
      }{48{q'}} + 
   \frac{11{{c}}^2{{q''}}^2{q^{(4)}}
      }{960{{q'}}^2} - 
   \frac{{{c}}^2{q'''}{q^{(4)}}
      }{80{q'}} + 
   \frac{131{c}{c'}{q^{(5)}}
      }{2880} - 
   \frac{{{c}}^2{q''}{q^{(5)}}
      }{240{q'}} + 
   \frac{{{c}}^2{q^{(6)}}}{180} \right)\,{{u_{x}}}^3{u_{xx}}
      \right.
      $$
$$\left.      
      + \left(\frac{7{c'}{c''}
      {q''}}{128} + 
   \frac{7{c}{c'''}{q''}
      }{384} - 
   \frac{7{{c'}}^2{{q''}}^2
      }{960{q'}} - 
   \frac{7{c}{c''}{{q''}}^2
      }{960{q'}} + 
   \frac{7{c}{c'}{{q''}}^3
      }{480{{q'}}^2} - 
   \frac{7{{c}}^2{{q''}}^4
      }{960{{q'}}^3} + 
   \frac{59{{c'}}^2{q'''}
      }{960} + 
   \frac{23{c}{c''}{q'''}
      }{320} - 
   \frac{{c}{c'}{q''}{q'''}
      }{30{q'}} 
      \right.\right.
      $$
      $$\left.\left. + 
   \frac{13{{c}}^2{{q''}}^2{q'''}
      }{640{{q'}}^2}- 
   \frac{3{{c}}^2{{q'''}}^2
      }{320{q'}} + 
   \frac{131{c}{c'}{q^{(4)}}
      }{1920} - 
   \frac{3{{c}}^2{q''}{q^{(4)}}
      }{320{q'}} + 
   \frac{31{{c}}^2{q^{(5)}}
      }{2880} \right)\,{u_{x}}{u_{xx}^2}
      \right.
      $$
      $$\left.+ 
   \left(\frac{3{c'}{c''}{q''}
      }{64} + 
   \frac{{c}{c'''}{q''}
      }{64} - 
   \frac{{{c'}}^2{{q''}}^2}{160{q'}} - 
   \frac{{c}{c''}{{q''}}^2
      }{160{q'}} + 
   \frac{{c}{c'}{{q''}}^3
      }{80{{q'}}^2} - 
   \frac{{{c}}^2{{q''}}^4}{160{{q'}}^3} + 
   \frac{47{{c'}}^2{q'''}}{960} + \frac{13{c}{c''}
      {q'''}}{240} 
      \right.\right.
      $$
      $$\left.\left. - 
   \frac{13{c}{c'}{q''}
      {q'''}}{480
      {q'}}+ \frac{{{c}}^2{{q''}}^2
      {q'''}}{60
      {{q'}}^2} - \frac{7{{c}}^2
      {{q'''}}^2}{960
      {q'}} + \frac{49{c}{c'}
      {q^{(4)}}}{960} - 
   \frac{7{{c}}^2{q''}{q^{(4)}}
      }{960{q'}} + 
   \frac{23{{c}}^2{q^{(5)}}}{2880}\right)\,{{u_{x}}}^2{u_{xxx}}
      \right.
      $$
      $$\left.
       + \left(\frac{5{{c'}}^2{q''}
      }{192} + 
   \frac{5{c}{c''}{q''}
      }{192} - 
   \frac{{c}{c'}{{q''}}^2
      }{96{q'}} + 
   \frac{{{c}}^2{{q''}}^3}{192{{q'}}^2} + 
   \frac{3{c}{c'}{q'''}
      }{64} - 
   \frac{{{c}}^2{q''}{q'''}
      }{96{q'}} + 
   \frac{{{c}}^2{q^{(4)}}}
    {96}\right)\,{u_{xx}}{u_{xxx}}
    \right.
    $$
    $$\left. + \left(\frac{{{c'}}^2{q''}
      }{64} + \frac{{c}{c''}
      {q''}}{64} - 
   \frac{{c}{c'}{{q''}}^2
      }{160{q'}} + 
   \frac{{{c}}^2{{q''}}^3
      }{320{{q'}}^2} + 
   \frac{9{c}{c'}{q'''}
      }{320} - 
   \frac{{{c}}^2{q''}{q'''}
      }{160{q'}} + 
   \frac{{{c}}^2{q^{(4)}}}
    {160}\right)\, {u_{x}} {u_{xxxx}}
    \right.
    $$
    $$\left.
     +\left( \frac{{c}{c'}{q''}
      }{192} - \frac{{{c}}^2{{q''}}^2
      }{960{q'}} + 
   \frac{{{c}}^2{q'''}}{480}\right)\,{u_{xxxxx}}\right]\delta(x-y)
$$}
\end{theorem}

To prove the Theorem one has to analyze the commutativity conditions
$$
{\mathcal E}\,  \left(\frac{\delta H_f}{\delta u(x)}\,  L \, \frac{\delta H_g}{\delta u(x)} \right) =0
$$
for arbitrary two functions $f(u)$, $g(u)$.
Here
$$
L= q\pal_x +\frac12 q' u_x -\frac{\epsilon^2 }8 c\, q' \pal_x^3 +\dots
$$
is the Hamiltonian differential operator associated with the second Hamiltonian structure.  To prove  validity of Jacobi identity one has to 
check that the $\epsilon$-terms in the second Hamiltonian structure can be eliminated by the quasitriviality transformation described in Section \ref{qua}. We will omit the calculations.

Observe that the family of bihamiltonian structures given in Theorem \ref{theorem3} depends on two arbitrary functions $c=c(u)$, $q=q(u)$, in agreement with the results of \cite{LZ1}.
It is understood that the Jacobi identity for the Poisson pencil holds true identically in $\lambda$  modulo terms of the order $O(\epsilon^6)$.

 \end{document}